\pdfoutput=1

\documentclass[letterpaper]{article}
\usepackage{iccc}
\usepackage{natbib}

\usepackage{subcaption}

\usepackage{times}
\usepackage{helvet}
\usepackage{courier}
\usepackage{appendix}
\usepackage{graphicx} %

\usepackage{hyperref}
\usepackage[all]{hypcap}

\usepackage{xcolor}
\definecolor{halfgray}{gray}{0.55}
\definecolor{webgreen}{rgb}{0,.5,0}
\definecolor{webbrown}{rgb}{.6,0,0}
\definecolor{webblue}{rgb}{0,0,0.930}
\definecolor{RoyalBlue}{cmyk}{1, 0.50, 0, 0}

\usepackage{hyperref}
\hypersetup{%
    colorlinks=true, linktocpage=true, pdfstartpage=1, pdfstartview=FitV,%
    breaklinks=true, pdfpagemode=UseNone, pageanchor=true, pdfpagemode=UseOutlines,%
    plainpages=false, bookmarksnumbered, bookmarksopen=true, bookmarksopenlevel=1,%
    hypertexnames=true, pdfhighlight=/O,%
    urlcolor=webblue, linkcolor=RoyalBlue, citecolor=webgreen, %
    } 
\usepackage{amsmath}
\usepackage{units}
\usepackage{siunitx}
\usepackage[capitalise]{cleveref}

\usepackage{booktabs}
\usepackage{array}
\newcolumntype{L}[1]{>{\raggedright\let\newline\\\arraybackslash\hspace{0pt}}m{#1}}
\newcolumntype{C}[1]{>{\centering\let\newline\\\arraybackslash\hspace{0pt}}m{#1}}
\newcolumntype{R}[1]{>{\raggedleft\let\newline\\\arraybackslash\hspace{0pt}}m{#1}}

\title{Exploring Conditioning for Generative Music Systems with Human-Interpretable~Controls}
\author{
Nicholas Meade\thanks{Equal contribution; ordering determined by coin toss.}\textsuperscript{,1},
Nicholas Barreyre\footnotemark[1]\textsuperscript{,1},
Scott C. Lowe\textsuperscript{1,2},
Sageev Oore\textsuperscript{1,2}\\
\textsuperscript{1}{Faculty of Computer Science, Dalhousie University, Halifax, NS, Canada}\\
\textsuperscript{2}{Vector Institute, Toronto, ON, Canada}\\
nicholas.meade@dal.ca,
nbarreyre@dal.ca,
scottclowe@gmail.com,
sageev@dal.ca\\
}

\begin{document} 
\maketitle
\begin{abstract}
Performance RNN is a machine-learning system designed primarily for the generation of solo piano performances using an event-based (rather than audio) representation.
More specifically, Performance RNN is a long short-term memory (LSTM) based recurrent neural network that models polyphonic music with expressive timing and dynamics \citep{Oore2018}.
The neural network uses a simple language model based on the Musical Instrument Digital Interface (MIDI) file format.
Performance RNN is trained on the \textit{e-Piano Junior Competition} Dataset \citep{piano-e-competition}, a collection of solo piano performances by expert pianists.
As an artistic tool, one of the limitations of the original model has been the lack of useable controls.
The standard form of Performance RNN can generate interesting pieces, but little control is provided over {\it what} specifically is generated.
This paper explores a set of conditioning-based controls used to influence the generation process.
\end{abstract}

\section{Introduction}

Computational, automated, and stochastic generation of music are pursuits of long-standing interest \citep{dice-games}; recently there has been an increasing body of research interest in these subfields~\citep{huang2018music,challenge-audio-2018-arxiv,dorien-survey,huang2018timbretron,payne-2019,music-vae}.

In a typical auto-regressive language model, the system generates a discrete probability distribution $P(\text{event}_0)$, samples from that distribution, and then uses its own sampled event history to condition the probability distribution over the next event to be predicted. An RNN model with a finite vocabulary is continually predicting $P(\text{event}_t | \text{event}_{i<t})$, where each $\text{event}_t$ is drawn from the vocabulary. PerformanceRNN, for example, is such a language model applied in a musical setting to generate expressive piano improvisations~\citep{Oore2018}.

One can adapt an auto-regressive language model so that its predictions are conditioned not only on the past events, but also on an externally-specified signal. For example, \citet{malik-ek} condition expressive generation on a score, and \citet{piano-genie} incorporate melodic pitch contours to provide very nice control over the generative mechanism.

In this work we explore further ways to provide the user with control over Performance RNN's generated musical output through a variety of conditioning signals, considered both individually and jointly. We begin by describing the data set used to train the system.

\begin{figure}[hbtp]
    \centering
    \includegraphics[width=1.0\columnwidth]{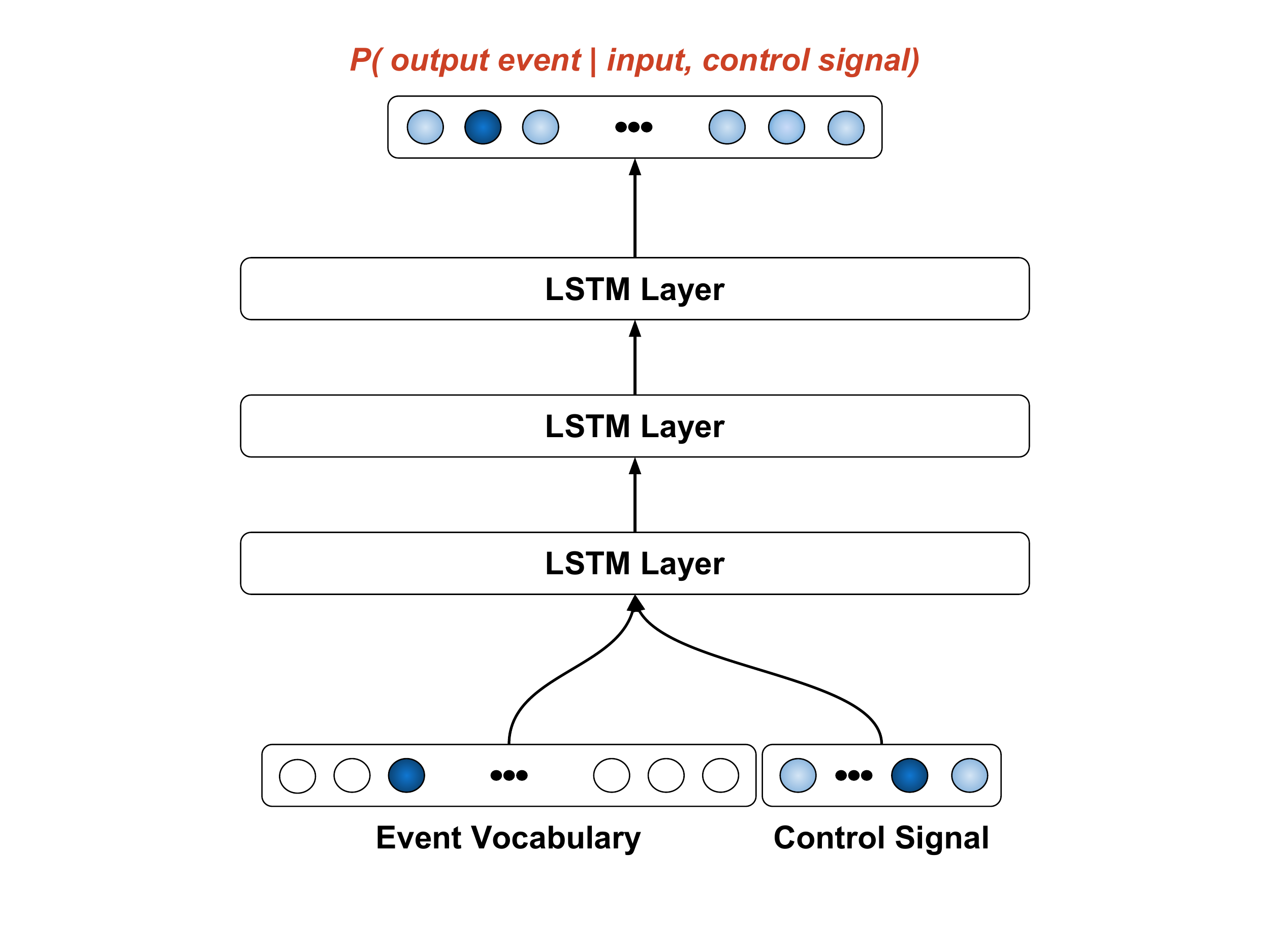}
    \caption{Performance RNN Architecture with Control Signals}
    \label{fig:architecture}
\end{figure}

The architecture we use to pass in control signals to Performance RNN is shown in \autoref{fig:architecture}.

\section{Dataset}

Similar to work by \citet{Oore2018}, we used the \textit{e-Piano Junior Competition} Dataset \citep{piano-e-competition} as training data for our models.
The raw MIDI files were converted to the Performance RNN representation, with a 388 word vocabulary consisting of 128 note-on, 128 note-off, 100 time-shift, and 32 velocity events.
Performances are modelled as sequences of these events, and are fed into the neural network using a one-hot vector encoding at each step.%
\footnote{The code both for the representation conversion and for the basic model were derived from the publicly available Magenta repository at \url{https://github.com/tensorflow/magenta/}.}

This dataset consists of 2750 performances by skilled pianists%
\footnote{More competition data has been released, allowing us to increase the size of the dataset from 1400 to 2750 performances.}.
The performances were split into training and validation partitions (90:10).
Additionally, we augmented the dataset with transpositions of up and down all intervals up to five or six semitones (spanning a full octave), and with temporal stretches/compression factors of 2.5\% and 5\%, increasing the number of training samples 35-fold.

We train the model on \SI{30}{s} segments from the MIDI performances using teacher forcing.

To achieve a conditional variant of Performance RNN, an additional feature vector is provided to the model along with each event, which we refer to as a \textit{control signal}.
At training time, the control signal provides additional information about the performance using metadata such as the composer of the piece.
When generating samples from the model, we can then constrain it to output a performance in the style of a single composer, for instance.

Various control signals were used, each described in more detail in the following sections.
Briefly, we had 3 sources for our control signals.
\begin{itemize}
\item Signals corresponding to the local statistics of the clip within the performance (note density, note velocity, and relative positioning within the piece).
\item Metadata directly available from the dataset (composer, and from this, attributes of the composer such as their year of birth).
\item Metadata extracted from the titles of the pieces (key, tempo and form).
\end{itemize}
For the first two sources, the metadata had complete coverage across the dataset.
However, only a minority of the titles indicated the key, tempo, or form of the piece.

\section{Control Signals}

In this section, we detail the control signals we explored, and the results for each.
The generated results mentioned below, along with additional samples, are all available at \href{https://doi.org/10.5281/zenodo.3277294}{doi:10.5281/zenodo.3277294}

\subsection{Composer-based Conditioning}

Though there are commonalities between composers, each composer has their own style of composition.
Being able to condition the model to generate music in the style of a particular composer would give us a useful control mechanism.

There are 114 different composers in the dataset, though some occur more frequently than other, less popular ones.
The distribution is approximately log-normal; around 53\% of the performances were pieces with one of the five most popular composers --- Chopin, Beethoven, Liszt, Schubert, and Bach.
From the model's perspective, most sequences in its world are composed by one of these five men, and an unconditioned model will hence generate music in one of their styles more often than not.
This highlights the utility of being able to select the composer whose style should be emulated.

We explored two types of control signals using the composer, either conditioning on the composer of the piece directly, or clustering the composers into groups and conditioning on the group.

\subsubsection{Individual Composers}

We used a 114-length feature vector representing the fraction of each composers' contribution to the training example.
Usually, this was a one-hot encoding; however, certain performances in the dataset were composed by multiple individuals.
These cases may arise, for instance, when one composer has written a piece for harpsichord and another composer has transcribed it into a piece for piano; in such cases all composers have influenced the final piece. We then assigned equal weighting in the control signal of $\nicefrac{1}{n_c}$, where $n_c$ is the number of composers.

We found that the model was able to express some of the stylistic differences between the composers (see \href{https://clyp.it/b3i2vbje}{Sample~10} and \href{https://clyp.it/igehzjff}{Sample~11}).
However, the distribution of the number of pieces per composer in the dataset has a long tail, and there were many composers for which only a single performance was available (around 29\% of composers only had a single piece in the dataset, meaning 1.2\% of the performances possess a unique composer).
For these one-off composers, it was not possible to both train a model conditioned on them and also to validate the model under that conditioning.
Furthermore, such a model could have tendency to overfit on composers associated with very few performances. We comment further on the significance of overfitting (or its insignificance) in the context of an artistic tool in the discussion section.

To quantify how well our composer-conditioned model captured the stylistic differences between composers, we surveyed five professional musicians with expertise in classical piano.
We selected five composers --- Bach, Beethoven, Chopin, Debussy, and Mozart --- and generated eight samples, each \SI{20}{s} in duration, for each composer. 
Each participant was given 10 samples (two from each composer) and tasked with rating the stylistic similarity of each clip to each of the five composers.
Scores were given from 1--5, where 1 denoted highly dissimilar and 5 was highly similar.
On average, participants scored the correct composer with a rating of 2.76 $\pm$ 0.18 (SEM) and incorrect composers with a rating of 1.95 $\pm$ 0.07 (SEM).

\subsubsection{Clustered Composers}

An expert pianist grouped 46 of the 114 composers into eight clusters (see \autoref{tab:composer-clusters}) based on style.
We used an input analogous to the individual composer control signal, except a 9 bit encoding was used to represent the distribution over the composer clusters instead.

\begin{table}[hbt]
\centering
\begin{tabular}{ll} \toprule
Cluster & Composers \\ \midrule
Cluster 1
 & Balakirev,\enskip Bartholdy,\enskip Bizet, \\
 & Brahms,\enskip Busoni,\enskip Chopin, \\
 & Grieg,\enskip Horowitz,\enskip Liszt, \\
 & Mendelssohn,\enskip Moszkowski,\enskip Paganini, \\
 & Saint-Saens,\enskip Schubert,\enskip Schumann, \\
 & Strauss,\enskip Tchaikovsky, Wagner \\
\rule{0pt}{2ex}Cluster 2
 & Beethoven \\
\rule{0pt}{2ex}Cluster 3
 & Bach,\enskip Handel,\enskip Purcell \\
\rule{0pt}{2ex}Cluster 4
 & Barber,\enskip Bartok,\enskip Hindemith, \\
 & Ligeti,\enskip Messiaen,\enskip Mussorgsky, \\
 & Myaskovsky,\enskip Prokofiev,\enskip Schnittke, \\
 & Schonberg,\enskip Shostakovich,\enskip Stravinsky \\
\rule{0pt}{2ex}Cluster 5
 & Debussy,\enskip Ravel \\
\rule{0pt}{2ex}Cluster 6
 & Clementi,\enskip Haydn,\enskip Mozart, \\
 & Pachelbel,\enskip Scarlatti \\
\rule{0pt}{2ex}Cluster 7
 & Rachmaninoff,\enskip  Scriabin \\
\rule{0pt}{2ex}Cluster 8
 & Gershwin,\enskip Kapustin \\
\rule{0pt}{2ex}Cluster 9
 & \textit{Everyone else} (unclustered) \\
\bottomrule
\end{tabular}
\caption{Composer clustering, constructed by hand by an expert pianist. From the composers, 46 were clustered into eight groups. The 68 unclustered composers were placed together in an additional, ninth group.} \label{tab:composer-clusters}
\end{table}

\subsubsection{Time-period Conditioning}

As with any art form, styles of composition evolve over time; many styles of classical music are associated with the period of history in which they originated and proliferated.

Unfortunately, the year of composition of each piece is not part of the metadata for this dataset.
As a proxy for the time period in which a piece was written, we used the year of birth of each composer.

We then grouped each performance by the century in which its composer was born: 1600, 1700, 1800, 1900, and \SI{2000}{CE}.
This was similar to the previously mentioned composer clustering, except that the binning of the composers was done based solely on their chronology without considering other factors.

In another variation, we normalized the year of birth of the composer to be a scalar in $[0,1]$ over the entire dataset, allowing us to interpolate and generate music conditioned on any year between roughly 1650 to \SI{2000}{CE}. 

In \href{https://clyp.it/hdoeacl2}{Sample~13} we can hear a clip conditioned to be similar to the 1600s.

\subsubsection{Composer Latitude, Longitude, and Birth Year}

Compositional styles are associated not only with a historical period, but also geographical regions.
The flow of musical knowledge and influence through both time and space motivated us to use a control signal based on the geographical locale of the composer, in addition to the time period in which it was written.

The city that was most associated with each composer provided us with geographic information about the music.
As above, we used the year of birth of the composer as a proxy for the date of authorship of each piece.
We used min-max normalization on the latitude, longitude, and year of birth, and represented this control signal as a vector of length three.
Thus, one component controlled movement from North to South, another from East to West, and a third, the year, spanned from \num{1653} to \SI{1972}{CE}.

For example, in \href{https://clyp.it/gtv4yaxp}{Sample~1} we can hear a sample generated conditionally on a city in Germany in 1685. Intuitively, we expect this to sound somewhat Bach-like. In \href{https://clyp.it/5d1ujskk}{Sample~2} and \href{https://clyp.it/2klwosxe}{Sample~3} we can hear two samples generated conditionally on Warsaw in 1810.
We would expect this to somewhat resemble Chopin. 
Overall, however, the year of birth tended to work more as expected than the location; it is likely that there was insufficient geographic variety in the data set, so that, for example, french impressionism did not sound particularly recognizable as such.
However, extrapolating to regions outside of the training distribution did occasionally produce quite interesting results, even if they did not make ``sense'' from the point of view of music history.

\subsubsection{Discussion}

Of the four mechanisms for conditioning based on the composer, we found that individual composer conditioning produced the most pleasing samples.
One reason why we think that composer clustering did not produce as high quality samples is that only 68 out of the 114 composers were placed into meaningful clusters, while the remaining ones were essentially counted as ``unknown''. As we discuss below, there is a potential problem with this.

\subsection{Title Keyword Conditioning}

Along with the MIDI files included as part of the e-Piano Competition data set, we also scraped the title of each performance from the web.

For some pieces, the title provided useful and interpretable information.
An example of such a title is \emph{Sonata in D Major, K. 576 (Complete) I. Allegro}, which we can easily determine is in the key D~Major, and marked tempo Allegro.
The signals that we extracted from the titles were key, tempo, and form of the piece.

\subsubsection{Encoding mechanism for partially labelled control signals}

As previously noted, some conditioning signals --- in particular, signals based on information present in the title --- were only available for a subset of the data.
For control signals which are always available to condition the model on during training, we supply the corresponding feature vector.
However, for control signals which are only partially observable, we must choose some default input to supply to the model in lieu of a feature vector.

Let us denote a multivariate control signal as $c=[c_1, c_2, \ldots]$.
When the control signal is unavailable, one option is to follow the methodology of PerformanceRNN's optional conditioning inputs.
With this method, when the control signal is unknown it is filled with zeros, and the conditioning signal is prepended by an additional bit, $c_0$, which indicates whether the control signal is provided.
For instance, the new conditioning input would be $[0, c_1, c_2, \ldots]$ when the control is available and $[1, 0, 0, \ldots]$ when it is not.
We experienced some difficulties using this encoding paradigm, particularly when control signals were very sparsely available --- for instance the tempo, which is only provided in the title of around 19\% of the pieces.
In these cases, the model would generate appropriate outputs only when the bit indicating the absence of the control signal was set to $1$.

We can provide some intuition for this failure mode as follows.
Let us consider the most extreme case: a control signal which is always absent from the dataset, so $c_0 = 1$ for all samples.
In this case, the conditioning signal is always set to $[1, 0, 0, \ldots]$; aside from the first bit (corresponding to absence of a control signal) the weights connected to the conditioning signal are all irrelevant to the model and will remain at their initialized values (unless weight-decay is present, in which case they decay to $0$).
Meanwhile, the weights connected to the first bit, $w_{c0}^{(j)}$, are each redundant with the respective neuron's bias term, $b^{(j)}$.
It is now important to the model only that the sum of these two parameters, $\hat{b}^{(j)} = b^{(j)} + w_{c0}^{(j)}$, is optimized, and the bias term $b^{(i)}$ itself may be very different to the bias term which would be learnt when training a model without the conditioning signal.
Without weight-decay, the difference between the two components will be the same after training as it was at initialization, since the parameters receive identical weight updates.
After training this model, inputs with $c_0 = 0$ will (clearly) not behave well since as the $w_{c0}^{(j)}$ terms are necessary to counteract the bias terms $b^{(j)}$ such that the effective bias, $\hat{b}^{(j)}$, is as optimised.
Ignoring the rest of the conditioning signal, if we were to include some inputs where $c_0 = 0$ during training, we would expect them to break the symmetry between $b^{(j)}$ and $w_{c0}^{(j)}$, and eventually $w_{c0}^{(j)} \to 0$.
However, the magnitude of the weights $b^{(j)}$ and $w_{c0}^{(j)}$ is of a similar order of magnitude to the activation of the neuron.
For a substantial fraction of the neurons where $w_{c0}^{(j)} > 0$, the neurons will have a negative preactivation whenever $c_0 = 0$, rendering them entirely inactive under a ReLU activation function.
Hence if the fraction of training samples where $c_0 = 0$ is insufficiently high, they will fail to break the symmetry before the model finds a local optima, at which many of the neurons are permanently dead for all samples with $c_0 = 0$.

We found better results could be achieved simply by omitting the $c_0$ bit from the conditioning vector during training.
This model can be conceptualised as learning a boosting procedure.
A ``baseline model'' is learnt which is used when the control signal is unknown and the conditioning signal is set to $c_i=0 \; \forall i>0$.
But when the conditioning signal is non-zero, the weights $w_{ci}^{(j)}$ are used to make fine-tuning improvements to the baseline model by increasing or decreasing the activation of each neuron in the first LSTM layer.
As a consequence, the residual error of the baseline model is reduced when the control signal is available.

For one-hot control signals, we also found good results by using a uniform distribution across $c_i$ when the true label was unknown (again, omitting a $c_0$ term).

\subsubsection{Major-Minor Conditioning}

The key of an excerpt of music is informative with regards to the pitches one would expect to dominate within the music, both in terms of number of occurrences and emphasis.
By extracting the key signatures that were present in performance titles, we were able to provide a control signal to the model corresponding to the key of the piece.

Ideally, we would like to condition on specific key signatures, such as A~minor and C~major.
However, only a small fraction of the 2750 performances had key signature information within their titles, so we grouped the keys into two clusters: major keys and minor keys.

There were numerous performances which contained neither ``major'' nor ``minor'' in their title. Because of this, our first implementation was a vector with three flags,
\begin{align*}
\text{major} \rightarrow& [1,0,0], \\
\text{minor} \rightarrow& [0,1,0], \\
\text{unknown} \rightarrow& [0,0,1].
\end{align*}
However, as described in the previous section on encoding partially labelled control signals, this was unsuccessful.
At generation time, the model was only able to generate music of comparable quality to the original, non-conditioned Performance RNN model if the control signal was set to $[0,0,1]$.
This was evidence for our aforementioned conjecture that an ``unknown flag'' is a poor way to represent sparsely annotated data.

\subsubsection{Tempo Keyword Conditioning}

We extracted relevant keywords from the titles pertaining to the tempo of the piece, and placed them into 5 tempo groups where tempos within the same group were more or less synonymous.
In addition, three expert musicians labelled the tempo of some additional pieces in our dataset.
The tempos which we considered were adagio, allegretto, allegro, andante, and presto.
Counts for each group are indicated in \autoref{tab:tempo-counts}.

\begin{table}[hbt]
\centering
\begin{tabular}{lr} \toprule
Tempo & Count \\ \midrule
Adagio & 57 \\
Andante & 131 \\
Allegretto & 160 \\
Allegro & 457 \\
Presto & 96 \\ \midrule  %
Total labelled & 901 \\
Unlabelled & 1849 \\ \bottomrule
\end{tabular}
\caption{Number of samples for each tempo, extracted from the titles of the pieces.} \label{tab:tempo-counts}
\end{table}

We attempted several different representations for tempo keywords.
First we tried a one-hot representation over the tempo groups with two additional components: one was a flag that indicating a mixed tempo and the other was a flag indicating whether the tempo was unknown.
Again, this representation's performance was underwhelming in practice, so the flags were removed and a zero vector was used for samples with unknown tempo. 

Results for the latter implementation were significantly better, especially in combination with (stochastic) beam search. 
Most convincingly, tempo controls can be interpolated (i.e. from fast to slow) at generation time and there is clear correspondence in the time of the music. For example, \href{https://clyp.it/un3tnsj4}{Sample~4} demonstrates an example of generation that starts at {\it{adagio}} (very slow) and is conditioned on {\it{presto}} (very fast) at the very end.
While the performances were not always pleasing to the ear, they followed the general trend of the given control.

\begin{figure}[hbtp]
    \centering
    \includegraphics[width=1.0\columnwidth]{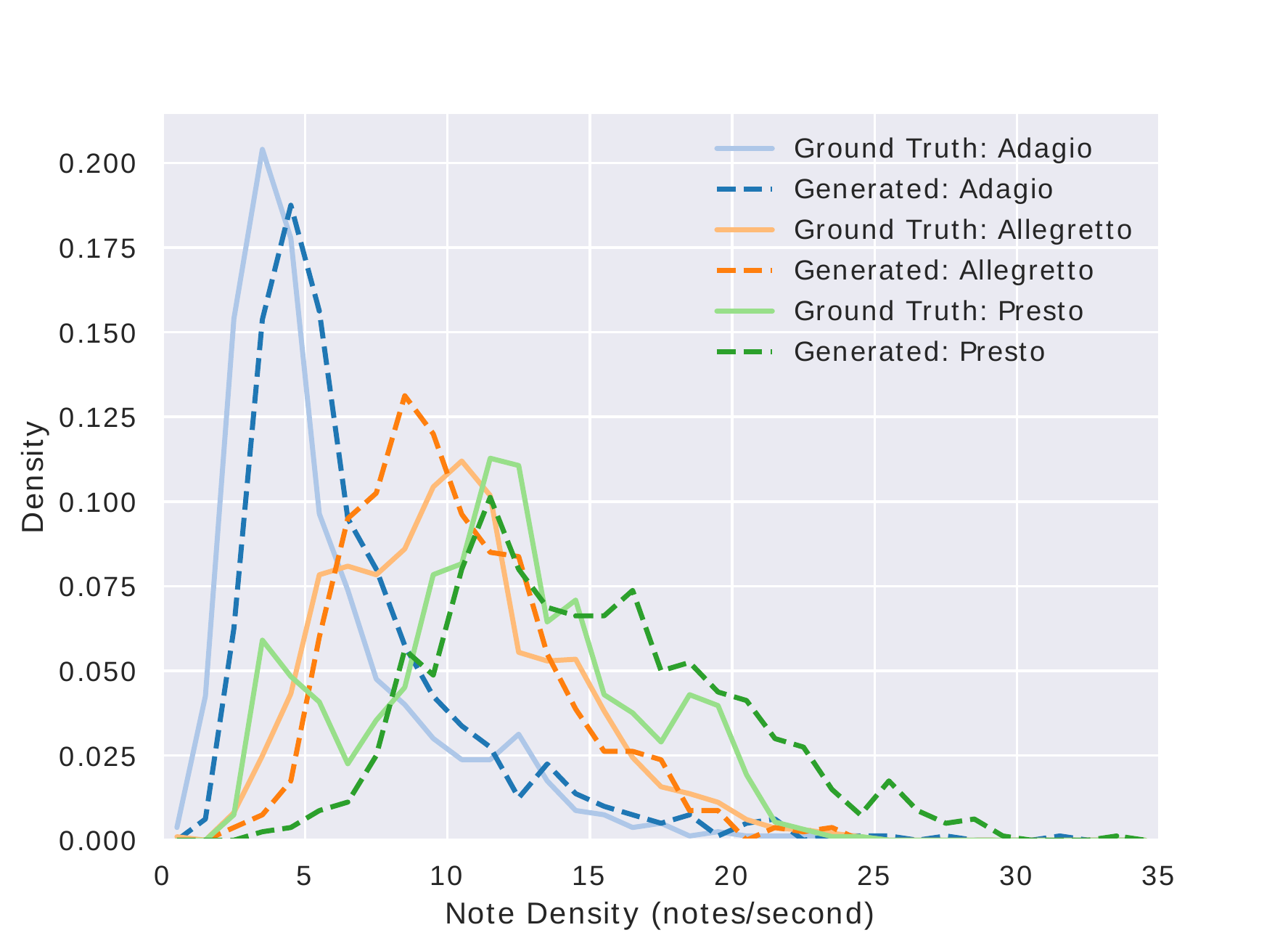}
    \caption{The distributions of note density for adagio, allegretto, and presto for both the generated samples and the ground-truth.}
    \label{fig:tempo-density}
\end{figure}

The tempo of a piece indicates its pace or speed.
Although tempo also conveys more nuanced information about the texture of the piece, broadly speaking, each tempo can be said to correspond to a certain number of beats per minute.

For evaluation purposes, as a proxy for the speed of a sample, we considered the note density --- the average number of notes played per second --- which can be evaluated computationally.
We investigated the distribution of note densities generated by our model when conditioned on each of the tempos that occurred in the training set.

We split each piece in the dataset into \SI{30}{s} segments, and computed the note density for each segment.
The note density was determined as the total number of note onset events in the segment, divided by its duration (\SI{30}{s}).
For each tempo conditioning value, we generated 800 samples each of \SI{45}{s} duration.
Each of these samples was cropped\footnote{We generated longer samples and then cropped them because it takes a short time for the network to settle after its initialisation.} to a final length of \SI{30}{s}, over which we again computed the note density.
We also generated 800 samples from an unconditional (``vanilla'') Performance RNN model trained without the tempo conditional signal.

The results (shown in \autoref{fig:tempo-density}) demonstrate the model learns the relationship between the tempo conditioning signal and the speed of the piece (in terms of notes per second).
Furthermore, the distribution of note densities are similar for the conditionally generated samples as to the ground-truth distributions from the source dataset.

\subsubsection{Form Keyword Conditioning}

Indicators of musical form were also present in the titles of many performances.
The forms we extracted, and the sample counts for each, are listed in \autoref{tab:form-counts}.
Similarly to the tempo keywords, we used this information to condition the model during training. 
\begin{table}[hbt]
\centering
\begin{tabular}{lr} \toprule
Form & Count \\ \midrule
Ballade & 48 \\
Dance & 6 \\
Espagnol & 5 \\
Etude & 397 \\
Fugue & 156 \\
Hungarian & 18 \\
Impromptu & 155 \\
Intermezzo & 13 \\
Mazurka & 7 \\
Polonaise & 31 \\
Prelude & 219 \\
Scherzo & 67 \\
Toccata & 32 \\
Variations & 106 \\
Waltz & 61 \\ \midrule  %
Total labelled & 1321 \\
Unlabelled & 1429 \\
\bottomrule
\end{tabular}
\caption{Number of samples for each form, extracted from the titles of the pieces.} \label{tab:form-counts}
\end{table}

Results were only obtained using the unknown-flag approach and the outcome was poor. Again, this provided evidence that such a paradigm does not work in practice. 
Further experiments are required to confirm our hypothesis that a zero-when-unknown encoding (or possibly some other alternative) will lead to better sounding results.

While the keywords in performance titles can be used to develop human-interpretable controls, they come with a great deal of noise.
For example, many of the MIDI files in the e-Piano Competition Dataset are actually recordings of multiple performances in sequence. What may be an accurate annotation for the first performance in the recording, is often completely inaccurate for the following pieces.

\subsection{Velocity Conditioning}

The velocity of a note strike describes how {\it hard} a note is played. Notes with high velocity are perceived as loud, while notes with low velocities are perceived as quiet. By providing a velocity-based conditioning signal to Performance RNN, we aim to be able to control the perceived volume of generated performances. We first point out that loud passages are not simply equivalent to quieter passages but with the volume turned up, just as yelling is not simply equivalent to a loud whisper. Nor does the content stay constant: the choice of notes, the phrasing, the articulation, may all likely be  distributed differently in loud passages when compared to quiet ones, just as what gets yelled is distributed differently, so to speak, from what gets whispered. Indeed, otherwise, increasing and decreasing all the velocities in a piece would have been another effective data augmentation technique.

The MIDI standard allows for velocities between 0 and 127, where 0 is the slowest possible velocity and 127 is the fastest.
The representation used by Performance RNN is coarser, quantized down by a factor of 4, yielding 32 velocity bins.
This provides a simpler input to the model, while still capturing most of the human-detectable difference between note velocities.
To construct our velocity conditioning signal, we further quantized these bins into three approximately equipopulated groups.
Roughly, these groups correspond to our perception of quiet, normal, and loud notes within a performance.
Notes with velocities from 0 to 14, 15 to 19, and 19 to 32 (in the Performance RNN representation) were placed into the quiet, normal, and loud bins respectively.

To construct a control signal for the training samples, we measured the distribution of note velocities across the three bins during each training sample (each sample having a duration of approximately \SI{30}{s}).
This conditioning signal for each sample was thus static through each training mini-batch.
At generation time, the human operator can select the velocity distribution to generate from, biasing the model towards either low, medium, or high velocities as they desire.
\href{https://clyp.it/hoag1g4h}{Sample~5}, \href{https://clyp.it/fohjkure}{Sample~6}, and \href{https://clyp.it/05cc5fys}{Sample~12} have been conditioned on velocity to begin quietly, grow loud, and then end quietly.

We observed that excerpts generated by the model tended to embody a velocity distribution very similar to the control signal.
To quantify how similar the distribution of velocities was, we computed the Kullback--Leibler (KL) divergence from the requested velocity distribution to the autoregressively-generated velocity distribution.
We uniformly sampled 3-bin velocity distributions $\vec{h} = [h_x, h_y, h_z]$ from the 2-d plane constrained by $h_x, h_y, h_z \in [0, 1]$ and $h_x + h_y + h_z = 1$.
For each sample $\vec{h}_{(i)}$, we autoregressively generated a single \SI{30}{s} audio clip using our model, trained as described above with a 3-bin velocity control signal, and measured the distribution of note velocities, $\vec{\hat{h}}_{(i)}$, in the associated MIDI file.
We measured the KL divergence from the distribution of velocities in the control signal to the generated distribution, $D_\text{KL}(\vec{h}_{(i)} \parallel \vec{\hat{h}}_{(i)})$, and repeated this process 100 times.
The median KL divergence was \SI{0.023}{bits}, with 95\% of samples falling in the range \SIrange{0.001}{0.168}{bits}.
This was statistically significantly smaller than our null hypothesis of independent distributions ($p < 0.001$).
To perform a statistical test on the median KL divergence, we independently sampled two 3-bin distributions $\vec{h}_{(j,1)}$ and $\vec{h}_{(j,2)}$ as described above and measured their KL divergence $D_\text{KL}(\vec{h}_{(j,1)} \parallel \vec{h}_{(j,1)})$.
This was repeated 100 times, and we took the median over these 100 repetitions to obtain a single estimate for $D_\text{KL}$ under the null hypothesis; this process was repeated 1000 times.
Under the null hypothesis, smallest observed value for the median $D_\text{KL}$ was \SI{0.335}{bits}.

We also constructed a temporally-dynamic velocity conditioning signal, using a \SI{5}{s} forward-looking window.
At each step of the model's training, the conditioning signal corresponded to the distribution of note velocities over the upcoming \SI{5}{s} worth of events.
However, this conditioning signal was very difficult to control when generating samples.
If a static velocity distribution is used throughout generation, the lack of dynamicism (which was present during training) confuses the model and causes it to generate a near ceaseless stream of note-onsets, refusing to produce either note-off or time-shift events.
We believe this failure mode is caused by the inconsistency between the history of notes input to the model (which are its own previous outputs), and the changes in the velocity signal (which does not change).
During training, the model can use the recent history of notes and the changes in the velocity signal to more accurately determine which velocities to produce; however when generating samples with a static velocity distribution this relationship breaks down.
While we could implement a dynamic conditioning signal during training, it is not clear that this would be successful if it was not also coupled to the notes generated by the model.

To increase the resolution of our control over the velocities, we also implemented a 5 class version of the static control signal.
Unlike the 3-bin variant, our 5-bins were not selected to be equipopulated.
Instead we hand-selected bin edges which allowed us to capture the extremes of the distribution, and in turn, a greater degree of control at generation time.
Our bins were $[0, 6]$, $[7, 14]$, $[15, 19]$, $[20, 23]$, and $[24, 31]$, determined in the Performance RNN quantization of velocity.

\subsection{Relative Position Conditioning}

A 30 second excerpt taken from a piece can vary greatly depending upon where in the piece it was taken from. Beginnings often differ significantly from endings, and climaxes are often distinguishable from both. With relative-position conditioning our aim is to be able to control roughly what part of a piece a generated performance sounds like. In other words, can we generate performances that sound like the beginning or end of a performance?

Each MIDI file used to train Performance RNN is augmented and split into a series of 30 second examples. With relative-position conditioning we provide an additional signal to the model indicating what position in the original source piece a particular example was taken from. For instance, an example with an initial conditioning signal of zero would begin at the start of a piece, while an example with a signal starting at 0.90 would begin 90\% through a performance. It is important to note that these signals increase within each example. As the example progresses through time, the signal increases proportionally.

During generation, the control signal is increased relative to the average performance length in the dataset.

\section{Joint Control Signals}

It is also possible to condition the model on multiple control signals simultaneously.
We explored the effect of conditioning the model of a pair of control signals at once, for several pairs of particular interest.

We did not attempt to train a model conditioned on more than two control signals simultaneously; if the amount of metadata provided to the model becomes too large, the model will receive enough information to identify exactly which piece the training sample is from, increasing the risk of overfitting.

\subsection{Relative-Position and Major/Minor Conditioning}

One problem faced while conditioning on major/minor was that the control signal, derived from the title, was not representative of the entire performance.
The key of the piece as stated in the title is often only accurate for the beginning (and end) of the performance.
For instance, a piece written in G Major may modulate to various other keys before it returns at the end to G Major.

To counteract this problem, we trained a model conditioned jointly on both the key (major/minor) as indicated in the title and the relative-position of the sample within the score.
This allowed us to generate samples conditioned on the beginning of pieces in either major or minor, where the key signature information would be most accurate. This did not work very consistently, however.

\subsection{Relative-Position and Composer}

We attempted to get our system to generate both beginnings and endings in the style of certain composers\footnote{We use the term ``style'' loosely here; we do not purport to be capturing the style of any of the composers at a deep level, just as many current image style transfer systems are not capturing the style of painters at a deep level.}. In \href{https://clyp.it/i3xa2qrg}{Sample~7} we can hear a clip generated to have characteristics of a Debussy-esque opening. Generally we found it quite hard to evaluate whether the outputs indeed sounded like openings or not. Endings were generally unsuccessful, although \href{https://clyp.it/ou5cun32}{Sample~8} demonstrates an attempt at a Bach ending where one can hear the final cadence a few notes near the very beginning, but then the system kept generating material after that.

\subsection{Tempo and Velocity Conditioning}
Using our results from the tempo and velocity conditioned models, we combined the zero-when-unknown vector representation with the five bin static velocity representation. 
Our results, especially when combined with beam search (as described below), clearly give the user control over tempo and velocity.
Nevertheless, the resulting samples often achieved their tempo and velocity settings differently than we expected.
In some cases, the generated samples contained a great deal of silence. 

In \href{https://clyp.it/zgpk22oe}{Sample~9}, we can hear a successful example of joint tempo and velocity control, where the clip was conditioned to start quietly (low velocity) and slowly ({\it{adagio}}), and then become loud (high velocity) and fast ({\it{presto}}). Notably, from roughly 0:06--0:09 the slow part contains a run of very fast notes, but the phrasing is such that it still has an unwaveringly slow feel, while the faster part never gets nearly as fast as that run, but has a significantly faster feel (although it is not quite as fast as a typical {\it{presto}}).

\section{Generation parameters}

\subsection{Beam search}

In the original Performance RNN, music was generated autoregressively, with each output conditioned on the previous output.
At each generation step, the output for that step is sampled from the distribution of possible outputs with probabilities equal to the likelihood values of each output as provided by the model.
The logits can optionally be rescaled with a temperature parameter before the sampling step; a high temperature increases the entropy of the distribution, whereas a temperature of 0 is equivalent to selecting the most likely output at each step.

A purely autoregressive model is a greedy search, selecting the output at each single step without consideration for the future generation steps.
However, sometimes it is better to select a less likely output for the current timestep in return for a payoff later of a more likely sequence overall.

One possible augmentation to this generation procedure is beam search.
With beam search, our goal is to generate a series of outputs which collectively have a high joint loglikelihood.
Throughout the beam search, we hold in memory $n_\text{beam}$ options (beams) simultaneously, along with the loglikelihood of the sequence for each beam.
For each beam, $f_\text{beam}$ (branch factor) copies are made and for each of these $n_\text{steps}$ outputs are autoregressively generated.
Of the $f_\text{beam} \cdot n_\text{beam}$ options, the $n_\text{beam}$ with the highest loglikelihood are retained.
This process is repeated until the length of the beams reaches the desired length, whereupon the beam with the highest loglikelihood is selected.

We found beam search was prone to generating outputs with locally low entropy, such as repeating the same note or same two notes throughout the piece, similar to using plain autoregression with a low temperature.
Intuitively, this is because generating a large number of samples from a distribution and then selecting the one with the maximum loglikelihood is equivalent to selecting the sample with highest loglikelihood.
To counteract this problem, we used a low branch factor of $f_\text{beam} = 2$ and a high $n_\text{steps} = 240\  \text{events}$, a duration equivalent to approximately 6 seconds of the performance.
We also chose $n_\text{beam} = 8$.
These parameters gave good results, but were not heavily optimised and we expect they could be improved upon.

Another variant of this is stochastic beam search, which selects which beams to retain with probabilities based on their loglikelihoods.
We also tried stochastic beam search (using a temperature of 1) with the same beam search parameters as above, and found this to give perceptually similar results.

\section{Discussion}

The generated results mentioned above, along with additional samples, are all available at \href{https://doi.org/10.5281/zenodo.3277294}{doi:10.5281/zenodo.3277294}.

Some conditioning paradigms give more fine-grained influence on the outputs of the model, such as the velocity distribution.
However, these are not necessarily easily interpretable by humans interfacing with the model.
Meanwhile, other controls such as the tempo are more easily understood but offer less nuanced control over the behaviour of the model.

Further work is required to determine the best representation for discrete and sparsely annotated control signals. Initial experiments were often framed from a probabilistic viewpoint; when annotations were certain, we used a value of one in its respective component. However this approach was combined with an ``unknown'' flag. 
While flags indicating the absence of a meaningful annotation are interpretable, they do not perform well in practice. 
Specifically, for tempo conditioning, we found that both uniformly distributing the input signal, and a vector of all zeros worked better than a flag approach when annotations were not available. 
Further experiments should include the expected value in place of an unknown annotation.  

There are numerous trade-offs that may be at play in the development and the functionality of machine learning (ML) based generative music systems. Some of these trade-offs arise from ML-related considerations, while others arise from human computer interaction (HCI) related considerations.
For example, a typical consideration in ML systems is the avoidance of overfitting; this is clearly understandable from a statistical perspective, and in our results we made efforts to present examples from models that we believe did not overfit. But from a generation perspective, where the goal is to provide artistic tools, some overfitting might not be a particularly negative quality, depending on its particular effects, and relative to other considerations.
For example, consider an auto-regressive generative model that is slightly overfit to certain training examples, i.e. musical passages, so that it occasionally recreates brief excerpts from those passages. This roughly corresponds to the notion of ``quoting'' other pieces and solos when improvising jazz solos. Artistically, that is not problematic at all: there are well-known solos which quote other well-known solos, and the downwards melodic run in Chopin's Fantasie-Impromptu is verbatim identical to a run at the end of Beethoven's Moonlight sonata.
If these are the effects of overfitting, then a bit of it is not necessarily negative.
Furthermore, if allowing for this can somehow provide an artistic tool with considerably more expressive user control, and indeed the user plans to be involved in the manipulation of the generated output, then relative to this criteria, the possibility of slight overfitting --- resulting in occasional quoting of the training material --- is an even lesser concern or possibly a benefit.

\section{Conclusion}
Interpretable controls for an LSTM-based RNN music generation system are possible.
In designing such a system, the representation of control signals appears to be an important factor, especially in dealing with sparsely annotated data. 
There is no question that we are able to control the output of the model at generation time, however, achieving the intended musical effect still remains a challenge.

\section{Acknowledgements}
Many thanks to Ian Simon, Sander Dieleman, Douglas Eck, and to the Magenta team at Google Brain.
We also thank Sidath Rankaduwa and Sonia Hellenbrand for assisting with labelling our dataset.
This work was carried out with the support of CIFAR, Natural Sciences and Engineering Research Council of Canada (NSERC) and \href{https://deepsense.ca}{DeepSense}.

\bibliographystyle{iccc}
\bibliography{sources}

\end{document}